 \documentstyle[twoside,fleqn,espcrc2]{article}

\newcommand{\AmS}{{\protect\the\textfont2
  A\kern-.1667em\lower.5ex\hbox{M}\kern-.125emS}}

\title{ Fermion induced  SU$(N)$ Yang-Mills Theory }

\author{Anna Hasenfratz \address{
Department of Physics, University of Colorado,
\\Boulder, Colorado 80309 USA}}

\begin{document}

\begin{abstract}
We investigate the gauge interaction induced by heavy fermions using both
dimensional and lattice regularization. We study the condition under
which heavy fermions induce a continuum gauge theory.

\end{abstract}

\maketitle


We study the low energy effects of heavy fermions in an $SU(N_c)$ gauge
theory. This talk summarizes the results presented in ref[1] and
extends that calculation for lattice regularization.

The effect of heavy matter fields at low energies
is expected to be
no more than an induced effective gauge coupling. This phenomena is well
known for lattice Wilson fermions where the fermion doublers induce a
shift in the gauge coupling of the order of $\Delta\beta\approx 0.3$ for
two flavors around $\beta=5.5$.
The idea that heavy fermions can induce continuum gauge fields (i.e. the
induced gauge coupling $\beta_{ind}=\infty$) has a
long history [2]. In
a recent paper it was suggested that one flavor of adjoint scalar
fields might induce a continuum gauge theory as well if the mass of the scalar
is
tuned appropriately [3]. It was argued furthermore that the scalar model can
be solved in the large $N_c$ limit.

 In this paper we calculate the gauge
coupling induced by heavy fermions in the fundamental representation and
investigate the condition under which the fermions induce a
continuum gauge theory. We argue that a necessary condition is that the
number of flavors $N_f > 11N_c/2$ but we cannot prove that it is a
sufficient condition as well. We consider the $N_f \to \infty$
sufficient condition here.

{\it A: Dimensional regularization. } Consider $N_f$ flavor of $SU(N_c)$
fermions with mass $M$ interacting with $SU(N_c)$ gauge fields $W_\mu$.
 The vacuum functional
         \begin{eqnarray}
       \int D \psi D \bar{\psi} DW_\mu \;\nonumber \\
        &\hspace{-1in} e^{ - \int_{x}
         \left [ \bar{\psi} \left ( \gamma^\mu \partial_\mu + M \right
         ) \psi- i \mu^{\epsilon/2} \;W^A_\mu \bar{\psi} \gamma^\mu
         T^A \psi \right ]}
\label{eq:vac}
         \end{eqnarray}
         is the standard, gauge invariant fermion-gluon interaction in
$n=4-\epsilon$ dimension, but the
         gluon part $\sim F^A_{\mu \nu} F^A_{\mu \nu}$ is missing.
(In  Eqn.~\ref{eq:vac} $T^A$, $A=1,...N_c^2-1$ are the $SU(N_c)$
generators.)
         Integrating over the fermions we get
         \begin{equation}
         \int DW_\mu \;\exp \left \{ - S(W_\mu) \right \},
         \end{equation}
         where the gauge field action $S(W_\mu)$ is a sum over
         one-loop graphs with $l$ gluon legs, $l= 2,3,\ldots$:
         \bigskip
         \begin{equation}
	 \vspace{.7cm}
\label{eq:seff}
         \end{equation}
         \bigskip
         The first term in  Eqn.~\ref{eq:seff}, giving the quadratic part of
the
         action has the form
         \begin{eqnarray}
        \frac{1}{g^2_0}\int_{p} \frac{1}{2}\; W^A_\mu(p) W^A_\nu (p)
         (p^2 \delta_{\mu\nu} - p_\mu p_\nu)\cdot \nonumber \\
         &\hspace{-2.1in} 6 \;\int^{1}_{0} dx\; x (1-x)
         \left [1 + x(1-x) \frac{p^2}{M^2} \right]^{-\epsilon/2},
\label{eq:quad}
         \end{eqnarray}
         with
         \begin{eqnarray}
         & &g_0  =  \bar{g}_0 \mu^{-\epsilon/2},
         \nonumber \\
         & &\frac{1}{\bar{g}^2_0}  =  N_f M^{-\epsilon} \;\frac{2}{3}\;
         \frac{1}{(4 \pi)^{n/2}}\; \Gamma \left (2- \frac{n}{2}\right ).
\label{eq:coupling}
         \end{eqnarray}
         For momenta $p^2/M^2 \ll 1$, Eqn.~\ref{eq:quad}  corresponds to
         $1/4{\bar g}^2_0 \int_x (\partial_\mu W^A_\nu (x) -
         \partial_\nu W^A_\mu (x))^2$. The graphs with 3 and 4 legs in
          Eqn.~\ref{eq:seff} make this expression gauge invariant, the square
         of the standard field strength tensor will enter. The bare
         gauge coupling $\bar{g}^2_0$ is proportional to $\epsilon$,
         goes to zero as the regularization is removed, as it should
         in an asymptotically free theory. The one-loop graphs with
         more than 4 legs in  Eqn.~\ref{eq:seff}
         are convergent and are suppressed
         by powers of  $1/M$ for momenta much below $M$.

When can we consider this model at low energies as a pure gauge theory?
A necessary condition is that the gluonic mass scale
$\Lambda_{\mbox{\tiny{MS}}}$ is
much lower than the fermionic mass scale $M$.  The leading term of the
effective action Eqn.~\ref{eq:seff}
is the usual $F_{\mu\nu}^AF_{\mu\nu}^A$ term with
bare coupling $1/{\bar g}_0^2$ given in  Eqn.~\ref{eq:coupling}.
If the higher order terms
in the effective action that are suppressed by powers of $1/M$ can be
neglected, we can express $\Lambda_{\mbox{\tiny{MS}}}$
in terms of the bare coupling
${\bar g^2_0}$
and the regularization parameter $\epsilon$. We will work out the
one-loop formula here. The two-loop derivation can be found in ref[1].

The one-loop definition of the $\Lambda$ parameter in the MS
renormalization scheme is
\begin{equation}
         \Lambda_{\mbox{\tiny{MS}}} = \mu
         \exp \left \{ - \frac{1}{2 \beta_0 g^2_{\mbox{\tiny{MS}}}
         (\mu)} \right \}.
         \end{equation}
where $\beta_0=11N_c/48\pi^2$ is the first (universal) coefficient of the
$\beta$-function.
Using the relation between the bare and renormalized $g^2_{\mbox{\tiny{MS}}}$
coupling
\begin{equation}
         \frac{1}{\bar{g}^2_0 \mu^{-\epsilon}} =
         \frac{2\beta_0}{\epsilon} + \frac{1}{g^2_{\mbox{\tiny{MS}}}
         (\mu)}
         \end{equation}
we obtain
\begin{equation}
 \Lambda_{\mbox{\tiny{MS}}} = \mu
         \exp \left \{\frac {1}{\epsilon}
         - \frac{1}{2 \beta_0 \bar{g}^2_0\mu^{-\epsilon}
         } \right \}.
         \end{equation}
If the gauge coupling is generated entirely by the fermions and has the
form given in Eqn.~\ref{eq:coupling}
\begin{equation}
 \Lambda_{\mbox{\tiny{MS}}} = \mu
         \exp \left \{\frac{1}{\epsilon}\left [ 1 -
\frac{N_f}{16\pi^2}\left ( \frac{M}{\mu}\right )
 ^{-\epsilon}\frac{2}{3 \beta_0 }
         \right ] \right \}.
         \end{equation}
The necessary condition $\Lambda_{\mbox{\tiny{MS}}}<<M$ can be satisfied as
$\epsilon \to 0$ only if
\begin{equation}
N_f > \frac{11}{2} N_c.
\label{eq:dim}
\end{equation}
The minimum number of flavors coincides with the value where the $\beta$
function of the gauge-fermion system changes sign and becomes
non-asymptotically free. The above derivation is valid only if the
higher order terms in the effective action can be neglected. Before
discussing this condition, let's derive the corresponding relations
using lattice regularization.

{\it B: Lattice regularization. }
Consider the action of $N_f$ Wilson fermions in the strong gauge
coupling limit
\begin{equation}
S=\frac{1}{2\kappa}\sum_{n,m} {\bar\psi}_n K_{nm}[U]\psi_m,
\end{equation}
where
\begin{eqnarray}
K_{nm}[U]=\delta_{nm}-\kappa\sum_\mu((r-\gamma_\mu)U_{n\mu}\delta_{n+\mu,m}
\nonumber \\
+(r+\gamma_\mu)U^{\dagger}_{n\mu}\delta_{n-\mu,m}).
\end{eqnarray}
$\kappa$ is related to the inverse fermion mass
\begin{equation}
\kappa= \frac{1}{2Ma+8r},
\end{equation}
where $a$ is the dimensional lattice spacing.
After integrating out the fermions we obtain the effective gauge action
\begin{eqnarray}
S_{eff}&=&-Tr ln K[U]\nonumber \\
&=&\sum_{\Gamma}\kappa^{l[\Gamma]}\frac{1}{l[\Gamma]}
Tr\left (\prod\Gamma(r\pm\gamma_\mu)\right )\;\cdot \nonumber \\
&&\left (TrU[\Gamma]+Tr U^{\dagger}[\Gamma]\right ),
\end{eqnarray}
where the sum is over all closed gauge loops.
Using the continuum representation of
 the gauge field  $U_{n\mu}=e^{igA_\mu(n)}$ one can express
$S_{eff}$ in terms of the continuum fields $A_{\mu}(n)$.
This calculation can be done using lattice perturbation theory or
following the method described in [3]. The coefficient of the
leading term $F_{\mu\nu}^AF_{\mu\nu}^A$ is given
by the four-dimensional lattice integral
\begin{eqnarray}
& & \hspace{-0.1in}\frac{1}{g^2_{latt}}=\nonumber \\
& & \hspace{-.1in} \frac{N_f}{4} \int d^4p Tr\left \{
Q(p_\mu)S(p)Q(p_\mu)\frac {\partial^2} {\partial p^2_\nu} S(p)\right \}
\label{eq:gl}
\end{eqnarray}
where $S(p)$ is the lattice fermion propagator
\begin{equation}
S^{-1}(p)=\frac{1}{2\kappa}-r\sum_\mu cos(p_\mu)-i\sum_\mu \gamma_\mu
sin(p_\mu)
\end{equation}
and $Q(p)$ is given by
\begin{equation}
Q(p_\mu)=i r sin(p_\mu) +\gamma_\mu cos(p_\mu).
\end{equation}
The integral reduces to the hopping parameter expansion result in the
$\kappa \to 0$ limit
\begin{eqnarray}
& &\frac {1}{g^2_{latt}}=4N_f \kappa^4 ,\quad r=1 ,\nonumber \\
& &\frac {1}{g^2_{latt}}=2N_f \kappa^4 ,\quad r=0 .
\end{eqnarray}
For small $Ma$ ($\kappa \to 0.125$) it has a logarithmic singularity
\begin{equation}
\frac {1}{g^2_{latt}}=\frac {N_f}{24\pi^2} ln \frac{\pi^2}{M^2a^2} ,\quad r=0 .
\label{eq:sm}
\end{equation}
One should note that for $Ma<<1$ the higher order tems in the effective
action are not necessarily suppressed and Eqn.~\ref{eq:sm} probably does
not give the induced gauge coupling correctly. Nevertheless,
assuming that the higher order terms in the effective action can be
neglected Eqn.~\ref{eq:sm} leads to
\begin{equation}
\frac {N_f}{16}>\frac {11}{2}N_c
\end{equation}
as a necessary condition for $\Lambda_{latt} << M$. If we observe that
the $r=0$ naive fermions describe 16 flavors of continuum fermions,
Eqn.~ \ref{eq:sm}
agrees with  the dimensional regularization result Eqn.~\ref{eq:dim}.

Now we turn our attention to the question of neglecting the higher order
terms in the effective action. Let's consider the general case where one
adds to the leading $F_{\mu\nu}^AF_{\mu\nu}^A$ term  some small gauge
         invariant perturbation which modifies the quadratic part and
         the vertices. The perturbation can be local, or nonlocal. We
         assume that the perturbation depends on some mass scale $M$ in
         such a way that for any set of fixed momenta flowing into the
         bare vertices or propagator the perturbation goes to zero as $M
        \rightarrow \infty$. Consider this modified model and take
         the limit $M \rightarrow \infty$ at {\em fixed} value of the
         regularization parameter. One expects that the effect of the
         perturbation disappears in this limit, the Green's functions
         remain unchanged. Consider any graph before and after the
         perturbation is introduced and take the difference. For any
         fixed set of internal and external momenta the integrand of
         the corresponding momentum integrals goes to zero as $M
         \rightarrow \infty$. If the regularization is such that the
 region of momentum integration is constrained  as is the case on the
lattice,
         then the integral itself goes to zero also, and the
         intuitive expectation is satisfied. The situation
         is less obvious if dimensional regularization is used and,
         actually, we are not able to present a formal proof.

In taking the $M\to\infty$ limit we have to consider the limit
where the bare charge is fixed at fixed regularization parameter,
i.e.
 \begin{eqnarray}
     && N_f \rightarrow \infty\;, \; M \rightarrow \infty, \nonumber \\
        &&N_f \cdot M^{-\epsilon} = c( \epsilon)\;, \;\mbox{fixed}
         \end{eqnarray}
for dimensional regularization with $\epsilon$ fixed
and
   \begin{eqnarray}
   && N_f \rightarrow \infty\;, \; M \rightarrow \infty,\nonumber \\
   && N_f/M^{4} = c^\prime(a)\;, \;\mbox{fixed}
         \end{eqnarray}
for lattice regularization with $\Lambda_{cut}=\pi/a$ fixed to obtain a
dimensionally/lattice regularized continuum  gauge model. As the
renormalization parameter is removed ($\epsilon \to 0$/$a \to 0$) the
bare coupling $g^2_0 \to 0 $ as it should in an asymptotically
free field theory. Clearly
the $N_f\to\infty$ condition is sufficient but perhaps not necessary to
obtain a continuum gauge theory. A lattice calculation might be able to
answer this question.

\section{ACKNOWLEDGEMENTS}
I am indebted to Peter Hasenfratz for the fruitful collaboration. I want
to thank T. DeGrand  for showing me a simpler way to derive Eqn.~\ref{eq:gl}.
This work was partially supported by NSF Grant PHY-9023257.

\newcommand{\PL}[3]{{Phys. Lett.} {\bf #1} {(19#2)} #3}
\newcommand{\PR}[3]{{Phys. Rev.} {\bf #1} {(19#2)}  #3}
\newcommand{\NP}[3]{{Nucl. Phys.} {\bf #1} {(19#2)} #3}
\newcommand{\PRL}[3]{{Phys. Rev. Lett.} {\bf #1} {(19#2)} #3}
\newcommand{\PREPC}[3]{{Phys. Rep.} {\bf #1} {(19#2)}  #3}
\newcommand{\ZPHYS}[3]{{Z. Phys.} {\bf #1} {(19#2)} #3}
\newcommand{\ANN}[3]{{Ann. Phys. (N.Y.)} {\bf #1} {(19#2)} #3}
\newcommand{\HELV}[3]{{Helv. Phys. Acta} {\bf #1} {(19#2)} #3}
\newcommand{\NC}[3]{{Nuovo Cim.} {\bf #1} {(19#2)} #3}
\newcommand{\CMP}[3]{{Comm. Math. Phys.} {\bf #1} {(19#2)} #3}
\newcommand{\REVMP}[3]{{Rev. Mod. Phys.} {\bf #1} {(19#2)} #3}
\newcommand{\ADD}[3]{{\hspace{.1truecm}}{\bf #1} {(19#2)} #3}
\newcommand{\PA}[3] {{Physica} {\bf #1} {(19#2)} #3}
\newcommand{\JE}[3] {{JETP} {\bf #1} {(19#2)} #3}
\newcommand{\FS}[3] {{Nucl. Phys.} {\bf #1}{[FS#2]} {(19#2)} #3}


\end{document}